\documentclass[12pt]{article}
\usepackage{graphics}
\begin{document}
\begin{center}
{The Self-energy of Nucleon for the Anomalous Magnetic Moment}
\end{center}
\begin{center}
{Susumu Kinpara}
\end{center}
\begin{center}
{\it National Institute of Radiological Sciences \\ Chiba 263-8555, Japan}
\end{center}
\begin{abstract}
The anomalous part of the magnetic moment of nucleon is studied.
The electromagnetic vertex of nucleon is calculated using the pseudovector coupling pion-nucleon interaction.
The self-energy of nucleon suggested in the previous study 
is applied to the internal line.
\end{abstract}
\section*{\normalsize{1 \quad Introduction}}
\hspace*{4.mm}
The interaction between two nucleons and their kinematic behavior are interesting and it is described by
the quantum field theoretical method like the meson-exchange model.
Regardless of the strength of the coupling parameters it is possible to resolve the properties of the two-body system 
by using the non-perturbative approach for the Bethe-Salpeter equation.
Besides the many-body system such as the finite nuclei 
also for the two-nucleon system the pion-exchange interaction is to be essential 
to interpret the experimental data.
Particularly the pseudovector coupling is not made clear about the role on the phenomena 
and the relation between the nuclear force and the quantum corrections at the present time. 
\\\hspace*{4.mm}
Because of the derivative on the field of pion the vertex part contains the variable on the four-momentum transfer
and then which makes difficult to obtain the convergent result unlike the pseudoscalar coupling.
While the divergences in the perturbative expansion are not removed within the usual procedure of the counter terms
the generalized relation remains and which is formed by the arbitrary number of the Heisenberg operators.
For the vertex function it is found to generate the non-perturbative terms in addition to the perturbative part.
Then for example the self-energy of the nucleon propagator results in the finite quantity which is applicable to
the calculation of the higher-order corrections.
\\\hspace*{4.mm}
One of the interesting subjects is the investigation of the electromagnetic properties of nucleon.
The form factor is associated with the magnetic moment of nucleon which is a basic property of fermion
and there are experimental data \cite{Exp} to be explained by the method of the field theory.
The anomalous part is larger than the Dirac part based on the point-like structure of nucleon.
Thus the degrees of freedom of pion have an effect on the electromagnetic form factor \cite{Chew}.
The lowest-order pion process is corrected by the nucleon propagator with the self-energy 
under the pseudovector coupling pion-nucleon interaction suggested in our previous study \cite{Kinpara}.
\section*{\normalsize{2 \quad The electromagnetic vertex function of nucleon}}
\section*{\small{2.1 \quad The relation between the vertex function and the self-energy}}
\hspace*{4.mm}
The charged pion interacts with photons as the quantum system under the lagrangian density given by
\\\begin{eqnarray}
&&L_{\pi-\gamma}(x) = -\frac{1}{4}F_{\mu\nu}(x)F^{\mu\nu}(x)\nonumber\\
&&+(\partial^\mu +i e A^\mu(x))\,\phi^{\ast}(x)(\partial_\mu -i e A_\mu(x))\,\phi(x) -m_\pi^2 \phi^{\ast}(x)\phi(x),
\end{eqnarray}
in which $\phi(x)$, $\phi^{\ast}(x)$ and $A_\mu(x)$ are the complex scalar field,
the conjugate field and the electromagnetic field respectively.
The strength of the electromagnetic field $F_{\mu\nu}$ is defined by $F_{\mu\nu}\equiv\partial_\mu A_\nu-\partial_\nu A_\mu$.
The operator $\phi(x)$ functions to destruct a negatively charged pion  
and create a positively charged pion (the electric charge $e>0$) and $m_\pi$ is the pion mass.
\\\hspace*{4.mm}
The interaction between $A_\mu(x)$ and $\phi(x)$ is obtained by the prescription of the minimal coupling 
$\partial_\mu\rightarrow\partial_\mu -i e A_\mu(x)$ which preserves the lagrangian invariant under the local gauge transformation that is $A_\mu(x)\rightarrow A_\mu(x)+\partial_\mu \alpha(x)$ 
and $\phi(x)\rightarrow {\rm exp}[i e \alpha(x)]\,\phi(x)$.
To proceed calculation the Lorentz condition $\partial^\mu A_\mu(x) = 0$ is imposed by determining the $\alpha(x)$ 
suitably and then the equation of motion for $A_\mu(x)$ is as follows
\\\begin{eqnarray}
\partial_\mu F^{\mu\nu}(x) = \partial_\mu \partial^\mu A^\nu (x) = J^\nu (x),
\end{eqnarray}
\begin{eqnarray}
J^\mu (x) \equiv e[\vec{\phi}(x)\times\partial^\mu\vec{\phi}(x)]_3
-e^2 A^\mu(x) (\phi_1^2(x)+\phi_2^2(x)).
\end{eqnarray}
Here the third component of the conserved isovector current $J^\mu (x)$ is expressed 
in terms of the real isovector fields $\vec{\phi}(x)=(\phi_1(x),\phi_2(x),\phi_3(x))$ 
instead of the fields $\phi(x)$ and $\phi^{\ast}(x)$
by using the relation between them $\phi(x)=(\phi_{1}(x)+i\,\phi_{2}(x))/\sqrt{2}$.
The subscript 3 in the first term tells that the neutral pion $\phi_{3}$ field is independent of $J^\mu (x)$.
The second term in Eq. (3) is not applied to the actual calculation below.
The vertex diagram needs another one of the photon propagators 
and so the numerical result would be suppressed by the factor ($\sim e^2$).
\\\hspace*{4.mm}
When the electromagnetic interaction for proton is taken into account the electromagnetic current 
in terms of the nucleon fields $\psi(x)$ and $\bar{\psi}(x)$
\begin{eqnarray}
J_{em}^\mu (x) \equiv e\,\bar{\psi}(x)\,\gamma^\mu \tau_{+} \psi(x) \qquad (\tau_{\pm}\equiv \frac{1\pm\tau_3}{2})
\end{eqnarray}
is added to the pion current $J^\mu (x)$ so that the source term in Eq. (2) is replaced as
$J^\mu (x) \rightarrow J^\mu (x)+J_{em}^\mu (x)$.
Here $\tau_3$ is the third component of the isospin matrix.
\\\hspace*{4.mm}
The photon-nucleon-nucleon three-point vertex part is used to investigate the electromagnetic properties of nucleon.
The vertex function $\it\Gamma$ is connected with the nucleon propagator $G$ and the expectation value
about the currents $J(z)$ and $J_{em}(z)$ by the following relation
\begin{eqnarray}
e \int d^4 x' d^4 y' \, G(x-x') \, \partial_z \cdot \Gamma(x'\,y'\,;\,z) \, G(y'-y) \nonumber\\
= e\,i\,\{\delta^4(z-x)-\delta^4(z-y)\}\tau_{+}\,G(x-y) \nonumber\\
-\langle\,{\rm T}[\partial_z \cdot (J(z) + J_{em}(z)) \, \psi(x)\, \bar{\psi}(y)]\,\rangle.
\end{eqnarray}
The second term in the right-hand side is left to proceed calculation of the anomalous part which is to be 
dropped by virtue of the current conservation.
Eq. (5) is converted to the analogous one in momentum space.
The relation between the vertex function $\Gamma(p,q)$ and the nucleon propagator $G(p)$ is given 
\begin{eqnarray}
(p-q)\cdot\Gamma(p,q) = \tau_{+}(G(p)^{-1}-G(q)^{-1})
\end{eqnarray}
as a function of the outgoing momentum $p$ and the incoming momentum $q$.
\\\hspace*{4.mm}
The vertex function $\Gamma(p,q)$ is divided into three parts 
\begin{eqnarray}
\Gamma(p,q) = \Gamma_{0}(p,q) + \Gamma_{\pi}(p,q) + \Gamma_{em}(p,q),
\end{eqnarray}
in which $\Gamma_{\pi}(p,q)$ and $\Gamma_{em}(p,q)$ are the counterparts of 
the terms of the currents $J(z)$ and $J_{em}(z)$ in Eq. (5) accordingly.
$\Gamma_0(p,q)$ satisfies Eq. (6). 
By the conservation $(p-q)\cdot\bar{u}(p) \Gamma_{0}(p,q) u(q) = 0$ 
using the Dirac spinors $u(q)$ and $\bar{u}(p)$ the component $\sim (p-q)$ is excluded from $\Gamma_{0}(p,q)$.
\\\hspace*{4.mm}
\section*{\small{2.2 \quad The anomalous magnetic moment by the pion current}}
\hspace*{4.mm}
For the calculation of the anomalous magnetic moment 
the pion current contribution $\Gamma_\pi^\mu(p,q)$ is the main part among the two components.
The isospin dependence is $\sim \tau_3$ 
and it is roughly the same value of the observed ratio between proton and neutron. 
By using the perturbative expansion 
the vertex function $\Gamma_\pi^\mu(p,q)$ is 
\begin{eqnarray}
&&\Gamma_\pi^\mu (p,q) = 2\,\tau_3\,(\frac{f_\pi}{m_\pi})^2 \int \frac{d^4 k}{(2 \pi)^4} \,(-2 k +p+q)^\mu 
\nonumber\\
&&\times\,\gamma_5\gamma\cdot(k-p)\,iG^{0}(k)\,\gamma_5\gamma\cdot(q-k)\,i \Delta^{0}(k-p)\,i \Delta^{0}(q-k)
\end{eqnarray}
in the lowest-order approximation. 
Here $G^{0}(k) = 1/(\gamma \cdot k-M+i\epsilon)$ and $\Delta^{0}(k) = 1/(k^2-m_\pi^2+i\epsilon)$ are the free propagators
of nucleon with the mass $M$ and pion.
\\\hspace*{4.mm}
It is noted that the pseudovector coupling interaction is connected to that of the pseudoscalar coupling and in the present case $\Gamma_\pi^\mu(p,q)$ is transferred to the pseudoscalar one by the equivalence relation 
$(\frac{f_\pi}{m_\pi})^2 = (\frac{g_\pi}{2 M})^2$ between these coupling constants 
and the replacement $\frac{\gamma \cdot (k-p)}{2 M} \rightarrow 1$, $\frac{\gamma \cdot (q-k)}{2 M} \rightarrow 1$.
Then the relation is such that the two interactions are agree with each other 
supposing the internal nucleon is restricted to the on-shell state ($k^2 = M^2$).
\\\hspace*{4.mm}
To perform the $k$-integral in Eq. (8) the Feynman formula is used and the variables are converted from $k$ to $k^\prime$
as $k^\prime = k-p x-q y$.
The denominator of the integrand becomes the simple form $({k^\prime}^2-M^2 \rho)^3$
where
\begin{eqnarray}
\rho \equiv (x+y-1)^2+\frac{m_\pi^2}{M^2}(x+y)-\frac{Q^2}{M^2} x y
\end{eqnarray}
and the four-momentum transfer $Q \equiv p-q$.
Then it is appropriate to use the method of the dimensional regularization integral.
Because of the sharp peak at ${k^\prime}^2 \sim M^2 \rho$ the quantity of the higher-order corrections would be treated approximately provided the form of it is analytic around the region.
\\\hspace*{4.mm}
In the case of the pseudovector coupling the estimate of the integration for $k$ is $\it\Gamma_\pi \sim k^{\rm 2}$ 
and then the convergence of the $k$-integration is not clear also about the $\sim(p+q)$ dependent term 
which is indispensable to calculate the anomalous part of the magnetic moment.
The nucleon current $\bar{u}(p)\Gamma(p,q)u(q)$ is expressed in the general form as
\begin{eqnarray}
\bar{u}(p)\Gamma^{\mu}(p,q)u(q) = \bar{u}(p)[\,\gamma^{\mu}F_1(Q^2)+\frac{i \sigma^{\mu\nu}Q_\nu}{2 M}F_2(Q^2)\,]u(q)
\end{eqnarray}
by means of the form factors $F_i(Q^2)\;(i=1,2)$ as a function of $Q^2$.
Here the relation of the Gordon decomposition is useful and which is
given by the form as $\bar{u}(p)[2M\gamma^\mu-(p^\mu+q^\mu)-i\sigma^{\mu \nu}Q_\nu]u(q)=0$.
\\\hspace*{4.mm}
Our interest is $F_2(Q^2)$ where the $k$-integral is done by the dimensional regularization method 
using two parameters $x$ and $y$ to treat the denominator of the integrand in Eq. (8) and it is
\begin{eqnarray}
F_2(Q^2) = \tau_3 \, (\frac{2 M f_\pi}{4 \pi m_\pi})^2 
\{\,(\frac{2}{\epsilon}-\gamma-{\rm log}\frac{M^2}{4 \pi \mu^2})I[f_1]+\sum_{i=2,3,4,5}I[f_i]\,\}, 
\end{eqnarray}
\begin{eqnarray}
I[f]\equiv\int_0^1 d x \int_0^{1-x} dy \, f(x,y), 
\end{eqnarray}
\begin{eqnarray}
f_1(x,y) \equiv -3 (x+y-1)(x+y-2)-(x+y)(x+y-3),  
\end{eqnarray}
\begin{eqnarray}
f_2(x,y) \equiv -f_1(x,y)\,{\rm log} \rho,  
\end{eqnarray}
\begin{eqnarray}
f_3(x,y) \equiv (x+y-1)(x+y-3),  
\end{eqnarray}
\begin{eqnarray}
f_4(x,y) \equiv \{\,(x+y-1)^4 - \frac{Q^2}{M^2} (x+y-1) (x+y-3) x y \,\}/\rho, 
\end{eqnarray}
\begin{eqnarray}
f_5(x,y) \equiv 4(x+y-1).
\end{eqnarray}
The infinitesimal quantity $\epsilon$ is stemmed from the shift of the space-time dimension as $4 \rightarrow 4-\epsilon$ and $\gamma$ 
is the Euler's constant ($\gamma=0.577\cdots$).
The parameter $\mu$ is introduced to set the mass dimension of the interacting lagrangian density to $4-\epsilon$.
Since $I[f_1]=0$ the parameters associated with the integral on $k$ are not significant to proceed the calculation. 
Then the result of the magnetic moment is free from the divergence contrary to the estimate mentioned above.
According to the dimensional regularization the additional term is required to calculate the constant term 
($\sim \epsilon^0$) correctly and in the present case it is given by $I[f_5]$.
It arises from the gamma matrix $\gamma_5$ which is commutable with the extra pieces of the gamma matrix
other than $\gamma_i$ ($i=0,1,2,3$) due to the shift of the dimension.
\section*{\small{2.3 \quad The anomalous magnetic moment by the vertex correction}}
\hspace*{4.mm}
For the electromagnetic form factor of nucleon the vertex correction by the pion propagator is important 
and as will be seen in the lowest-order calculation 
the magnitude of the numerical value is as large as that of the pion current apart from the isospin matrix.
In fact the observed $\sim \tau_3$ dependence of the anomalous part does not necessarily
imply that the process is minor taking into account the non-perturbative effect of the self-energy of nucleon.
\\\hspace*{4.mm}
The vertex function of the electromagnetic current $\Gamma_{em}^\mu (p,q)$ is 
\begin{eqnarray}
&&\Gamma_{em}^\mu (p,q) = -\,(1+\tau_{-})\,(\frac{f_\pi}{m_\pi})^2 \int \frac{d^4 k}{(2 \pi)^4} \, i \Delta^{0}(k) 
\nonumber\\
&&\times\,\gamma_5\gamma\cdot k\,iG^{0}(p-k)\,\gamma^\mu \,iG^{0}(-k+q)\,\gamma_5\gamma\cdot k
\end{eqnarray}
in the lowest-order approximation. 
As well as the form factor of the pion current the equivalence relation exists 
if the two internal nucleon propagators are at the on-shell states ($(p-k)^2=M^2$ and $(q-k)^2=M^2$).
\\\hspace*{4.mm}
The $k$-integral is done same as the $\Gamma_{\pi}^\mu (p,q)$ case using $\rho^\prime$ given by 
\begin{eqnarray}
\rho^\prime \equiv (x+y)^2+\frac{m_\pi^2}{M^2}(1-x-y)-\frac{Q^2}{M^2} x y
\end{eqnarray}
for the denominator $({k^\prime}^2-M^2 \rho^\prime)^3$.
When $Q^2=0$ both $\rho^\prime$ and $\rho$ are positive definite and it suffices 
to do the integrals for the analytic functions.
They are transferred to each other by the interchange $x+y \leftrightarrow 1-x-y$.
The form factor of the nucleon current corrected by the lowest-order pion process is 
\begin{eqnarray}
F_2(Q^2) = (1+\tau_{-}) \, (\frac{2 M f_\pi}{4 \pi m_\pi})^2 
\{\,(\frac{2}{\epsilon}-\gamma-{\rm log}\frac{M^2}{4 \pi \mu^2})I[f_6]+\sum_{i=7,8,9}I[f_i]\,\}, 
\end{eqnarray}
\begin{eqnarray}
f_6(x,y) \equiv 3 (x+y) - 2,  
\end{eqnarray}
\begin{eqnarray}
f_7(x,y) \equiv -f_6(x,y)\,{\rm log} \rho^\prime,  
\end{eqnarray}
\begin{eqnarray}
f_8(x,y) \equiv -x-y,  
\end{eqnarray}
\begin{eqnarray}
f_9(x,y) \equiv \{\, -(x+y)^3-\frac{Q^2}{M^2}(2-x-y)x y  \,\}/{\rho^\prime}.
\end{eqnarray}
In Eq. (18) two additional terms on the commutability of $\gamma_5$ cansel with each other 
and do not contribute to the coefficient of the term $\sim p+q$.
So there is not the effect on the magnetic moment.
The divergent term vanishes by $I[f_6]=0$ also in the part of the nucleon current.
\section*{\normalsize{3 \quad The numerical results and the effect of the self-energy}}
\hspace*{4.mm}
The numerical calculation is done by using the set of the parameters $M=939$ MeV, $m_\pi=139.6$ MeV and $f_\pi=1.0$.
At $Q^2=0$ the values of the integrals $I[f_i]\;(i=2,3,4)$ are $I[f_2]=0.896$, $I[f_3]=0.417$, $I[f_4]=0.079$ 
and $I[f_5]=-2/3$.
Adding them the result is $F_2(0) = a\,\tau_3$ with $a=0.831$ 
which constructs the electromagnetic pion current contribution for the anomalous magnetic moment in the lowest-order.
The part of the lowest-order vertex correction for the nucleon current is determined similarly
by the integrals $I[f_i]\;(i=7,8,9)$ with the values of $I[f_7]=-0.288$, $I[f_8]=-1/3$ and $I[f_9]=-0.326$. 
It results in $F_2(0) = b\,(1+\tau_{-})$ with $b=-1.086$. 
\\\hspace*{4.mm}
In order to compare the result with the experimental value the form factor at $Q^2 = 0$ is expressed as
\begin{eqnarray}
F^{exp}_2(0) = \kappa_{p} \,\tau_{+} + \kappa_{n} \,\tau_{-} 
= z\, \tau_{+} + a^\prime(z)\,\tau_3 + b^\prime(z)\,(1+\tau_{-})
\end{eqnarray}
\begin{eqnarray}
a^\prime(z) \equiv \frac{-2 z +2 \kappa_{p}-\kappa_{n}}{3}
\end{eqnarray}
\begin{eqnarray}
b^\prime(z) \equiv \frac{-z +\kappa_{p}+\kappa_{n}}{3}
\end{eqnarray}
incorporating the terms of the self-energy and the vertex correction by the pion propagator too.
The parameter $z$ is ascribed to the self-energy $\Sigma(p)$ of the nucleon
given by a series of $\gamma \cdot p - M$.
The experimental values of the proton and the neutron anomalous magnetic moments are 
$\kappa_{p}=1.79$ and $\kappa_{n}=-1.91$ in units of the nuclear magneton ($\mu_N$) respectively \cite{Exp}.
The coefficients $a^\prime(z)$ and $b^\prime(z)$ are linearly dependent on $z$ 
and the convergent results of the perturbative
expansion may give the information on the magnitude of the self-energy.
It is obtained by means of the non-perturbative part of the vertex function.
\\\hspace*{4.mm}
In Fig. 1 the linear functions $a^\prime(z)$ and $b^\prime(z)$ on $z$ are drawn along with the results of the calculation
of the coefficients $a$ and $b$ as the horizontal lines.
When $z=0$ the $a$ and $b$ are different largely from the experimentally determined values $a'(0)$ and $b'(0)$ 
for lack of the self-energy in Eq. (6).
This situation is changed by moving $z$ to $z=2$ at which the $a/a'(2)$ and $b/b'(2)$ are comparable
such as $a/a'(2) \sim b/b'(2) \sim 1.6$ and the relation disappears again as $z$ continues to increase.
It indicates that the present calculation would be improved by the higher-order corrections such as to
reduce the value of the coupling constant $f_\pi$ in $a$ and $b$ ($\sim f_\pi^2$) effectively.
\\\hspace*{4.mm}
Another kind of the higher-order correction is the inclusion of the self-energy for the nucleon propagator
in the off-shell state.
When there is not the pion interaction the photon-nucleon-nucleon three-point vertex function is in the perturbative
expansion prescribed by the Feynman rule.
The magnitude of the fine structure constant as the parameter for the expansion makes us understand 
the iterative character of the method.
Taking into account the pseudovector coupling interaction of pion this situation is changed 
because the parameter has the dimension of [MeV]${}^{-2}$ 
which implies the expansion series are associated with divergences worse than that in the quantum electrodynamics.
\\\hspace*{4.mm}
The non-perturbative treatment of the pion-nucleon-nucleon three-point vertex function provides the non-perturbative term 
which is excluded from the rule of the perturbative expansion.
Consequently the proper vertex function is modified in comparison to the one without the term.
As has been seen in our previous study the effects on the propagators are not negligible.
The self-energy of the nucleon propagator becomes 
\begin{eqnarray}
\Sigma(k) = \frac{M\,c^2}{M^2+m_\pi^2-\frac{M c}{2}-\frac{c^2}{2}}=\frac{M}{M^2+m_\pi^2}c^{2} + O(c^3)
\end{eqnarray}
by the lowest-order approximation for the perturbative part of the vertex function \cite{Kinpara}
and which is expanded in the series of $c \equiv \gamma \cdot k - M $.
\\\hspace*{4.mm}
We do not use Eq. (28) directly for the calculation of the higher-order corrections of the anomalous magnetic moment.
Alternatively the off-shell state of the propagator is considered to have the effective mass $M^\prime$ defined by
$M^\prime \equiv M + \langle\Sigma(k)\rangle$, in which the bracket means $\Sigma(k)$ is
replaced by an approximate value independent of $k$.
The approximation of the internal nucleon state achieves to include the effect of the higher-order correction
preserving the simple form of the free propagator.
\\\hspace*{4.mm}
The four-momentum $k$ of the internal off-shell nucleon is the variable of the integral and permitted to take any values.
Determining the value of the self-energy $\langle\Sigma(k)\rangle$ 
we replace as $\gamma\cdot k \rightarrow t$ and $k^2 \rightarrow t^2$ in Eq. (28) 
using a parameter $t$ with the dimension of mass.
At the region $0 \leq t \leq 2 M$ suitable for the present study 
the effective mass $M^\prime$ is larger than $M$.
\\\hspace*{4.mm}
There exists a way to determine the specific value of $M^\prime$.
The inverse of the exact propagator $G^{-1}(k)$ has the zero at the parameter $t = M+\Sigma(k)$ 
where $\gamma \cdot k \rightarrow M+\Sigma$ is done in $\Sigma(k)$.
Then the self-consistent equation on $\it\Sigma$ is derived by means of the form of $\it\Sigma$ in Eq. (28).
The positive solution is $\Sigma = M (-3+\sqrt{17+8 m_\pi^2/M^2})/2\sim 0.56 M\sim 530\,{\rm MeV}$ roughly 
as large as the $\sigma$ meson mass.
Consequently the coefficients on the pion current $a$ and the nucleon current $b$ are $a = 0.758$ and $b = -1.077$ 
respectively and the ratios become $a/a'(2) \sim b/b'(2) \sim 1.5$.
Improvement is observed however it is not considerable.
While it is not possible to explain the data satisfactorily
the approximation is related to the structure of $G(k)$.
\section*{\normalsize{4 \quad Concluding remarks }}
\hspace*{4.mm}
The magnitude of the anomalous part is larger than the Dirac part in the magnetic moment of nucleon.
The dependence on the isospin makes us convinced that the virtual processes of the charged pions actually take place
and which is observed by the electromagnetic field.
The pseudovector coupling interaction is useful to describe it quantitatively.
The result is free from divergences within the lowest-order approximation.
Inclusion of the self-energy of nucleon is necessary and its effect has been investigated 
in the present study.
The approximate propagator with the effective mass is tractable since it is connected to the lowest-order calculation.
The change of the nucleon mass may reflect the mixing of the continuum state 
beyond the one-particle state in the exact propagator.

\newpage
\begin{figure}
\begin{center}
\scalebox{0.5}{\includegraphics{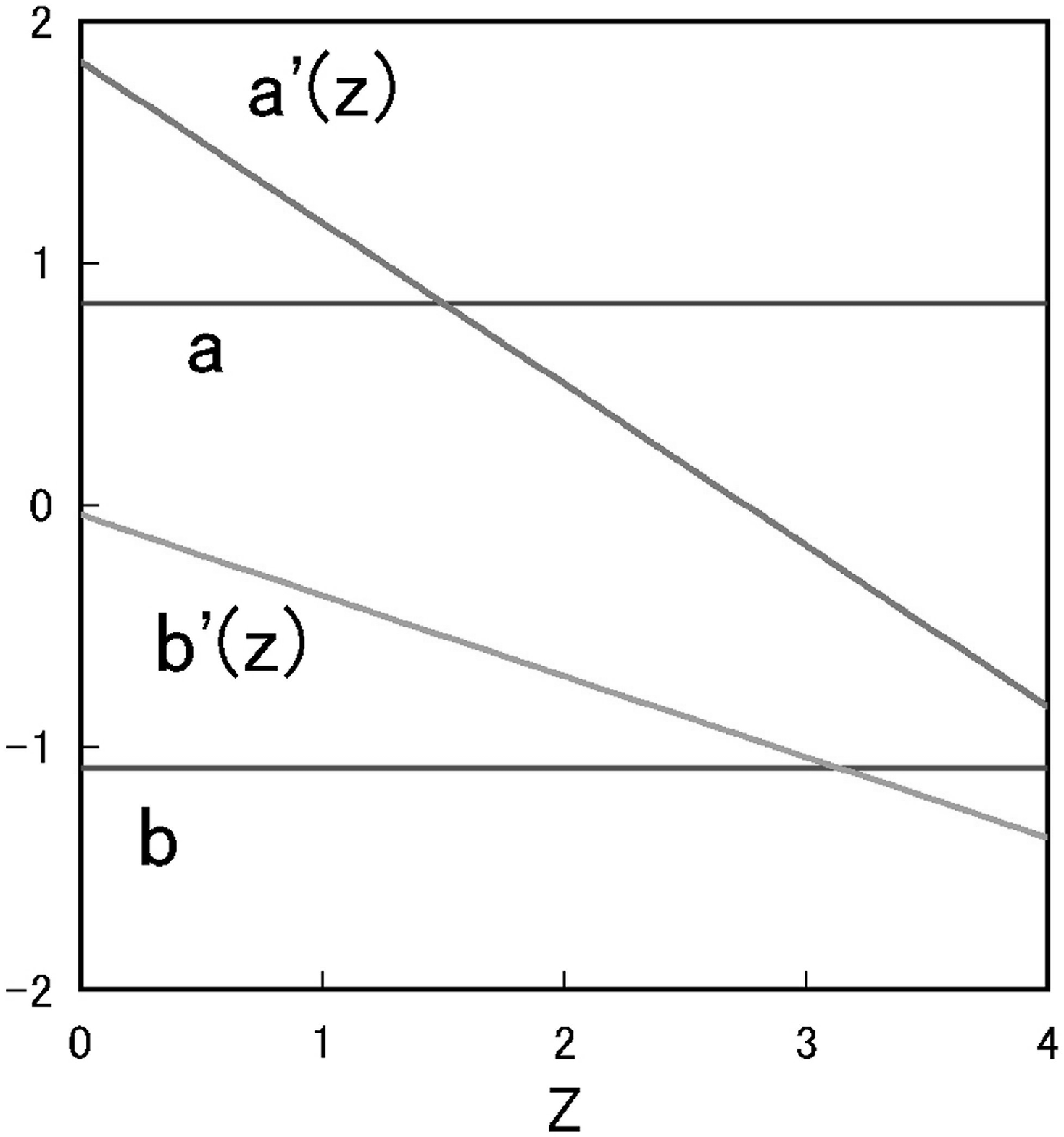}}
\caption{
The coefficients $a^\prime(z)$ and $b^\prime(z)$ are shown as the linear function of $z$.
The results of the calculations $a$ and $b$ are shown by the horizontal lines.  
}
\end{center}
\end{figure}
\end{document}